\documentstyle[12pt]{article}

\newcommand{\jx}{j\,(x)}

\newcommand{\jsx}{j^*\,(x)}
\newcommand{\jy}{j\,(y)}

\newcommand{\jxh}{j\,\left({\frac x 2}\right)}
\newcommand{\jmxh}{j\,\left( - {\frac x 2}\right)}
\newcommand{\kjxy}{[j\,(x), j\,(y)]}

\newcommand{\xmys}{{(x - y)}^2}
\newcommand{\xzmyzs}{{(x_0 - y_0)}^2}
\newcommand{\xtmyts}{{(x_3 - y_3)}^2}
\newcommand{\vq}{\vec{q}}

\newcommand{\xz}{x_0}
\newcommand{\xo}{x_1}

\newcommand{\xt}{x_3}

\newcommand{\yz}{y_0 }

\newcommand{\qz}{q_0}
\newcommand{\qzpr}{q'_0}

\newcommand{\dx}{d\,x}

\newcommand{\dfx}{d^4\,x}

\newcommand{\ez}{E_0 }
\newcommand{\ep}{E^{'} }
\newcommand{\eimej}{(E_i - E_j) }

\newcommand{\emei}{(E - E_i) }
\newcommand{\eime}{(E_i - E) }

\newcommand{\de}{\delta }

\newcommand{\vf}{\varphi }
\newcommand{\vfx}{\varphi\,(x) }
\newcommand{\vfy}{\varphi\,(y) }
\newcommand{\vfxh}{\varphi\,\left({\frac x 2}\right) }
\newcommand{\vfmxh}{\varphi\,\left(- {\frac x 2}\right) }

\newcommand{\eps}{\varepsilon }
\newcommand{\bfe}{\Phi\,(E) }
\newcommand{\bfee}{\Phi_{\varepsilon}\,(E) }
\newcommand{\tbfe}{\tilde \Phi\,(E) }
\newcommand{\tbfee}{\tilde \Phi_{\varepsilon}\,(E) }

\newcommand{\teod}{\theta_{12} }
\newcommand{\tedo}{\theta_{21} }
\newcommand{\tezi}{\theta_{0i} }

\newcommand{\fevq}{F\,(E, \vec{q}) }

\newcommand{\frevq}{F^{ret}\,(E, \vec{q}) }
\newcommand{\fraevq}{F^{\stackrel {ret}{adv}}\,(E, \vec{q}) }
\newcommand{\fpevq}{F_+\,(E, \vec{q}) }
\newcommand{\fmevq}{F_-\,(E, \vec{q}) }
\newcommand{\fpmevq}{F_{\pm}\,(E, \vec{q}) }
\newcommand{\tfevq}{\tilde F\,(E, \vec{q}) }

\newcommand{\fx}{F\,(x) }
\newcommand{\fax}{F^{adv}\,(x) }
\newcommand{\frx}{F^{ret}\,(x) }
\newcommand{\frmx}{F^{ret}\,(- x) }
\newcommand{\frax}{F^{\stackrel {ret}{adv}}\,(x) }
\newcommand{\fpx}{F_+\,(x) }
\newcommand{\fmx}{F_-\,(x) }
\newcommand{\fpmx}{F_{\pm}\,(x) }
\newcommand{\fpxm}{F_+\,(- x) }
\newcommand{\fxm}{F\,(- x) }

\newcommand{\fxc}{F^{c}\,(x) }
\newcommand{\fe}{F\,(E) }

\newcommand{\fepr}{F\,(E^{'}) }
\newcommand{\fee}{F_{\varepsilon}\,(E) }
\newcommand{\tfee}{\tilde F_{\varepsilon}\,(E) }
\newcommand{\feepr}{F_{\varepsilon}\,(E^{'}) }

\newcommand{\femepiz}{F_{\varepsilon}\,(- E + i\,0)}
\newcommand{\fesepiz}{F^{*}_{\varepsilon}\,(E + i\,0)}
\newcommand{\tfeemiz}{\tilde F_{\varepsilon}\,(E - i\,0)}
\newcommand{\kse}{\xi\,(E) }

\newcommand{\fsepiz}{F^{*}\,(E + i\,0)}
\newcommand{\tfemiz}{\tilde F\,(E - i\,0)}

\newcommand{\feei}{F_{\varepsilon}\,(E_i)}
\newcommand{\psee}{{\psi}_{\varepsilon}\,(E)}
\newcommand{\tpsee}{\tilde {\psi}_{\varepsilon}\,(E)}

\newcommand{\pseep}{{\psi}_{\varepsilon}\,(E^{'})}
\newcommand{\tpseep}{\tilde {\psi}_{\varepsilon}\,(E^{'})}
\newcommand{\pser}{{\psi}_{\varepsilon}\,(R\,e^{i\,\varphi})}

\newcommand{\taxz}{\tau\,(x_0)}
\newcommand{\tamxz}{\tau\,( - x_0)}

\newcommand{\exs}{e^{i\,(E\,x_0 - \vec{q}\,\cdot\,\vec{x})}}
\newcommand{\exiexz}{e^{i\,E\,x_0}}
\newcommand{\exieprxz}{e^{i\,E'\,x_0}}

\newcommand{\exxt}{{e^{- i\,e_3\,x_3\,\sqrt{E^2 - E_0^2}}}}
\newcommand{\exq}{e^{- i\,(q_1\,x_1 + q_2\,x_2)}}
\newcommand{\cosxt}{{\cos{(e_3\,x_3\,\sqrt{E^2 - E_0^2})}}}
\newcommand{\cosqx}{{\cos(\vec{q}\,\cdot\,\vec{x})}}
\newcommand{\sqez}{{\sqrt{E^2 - E_0^2}}}

\newcommand{\expa}{e^{i\,(p^{'}  - p)\,a}}

\newcommand{\dedevfx}{\frac{\de}{\de\,\vfx} }

\newcommand{\dedevfmxh}{\frac{\de}{\de\,\vfmxh} }
\newcommand{\desdevfx}{\frac{\de\,S}{\de\,\vfx} }
\newcommand{\desdevfxh}{\frac{\de\,S}{\de\,\vfxh} }
\newcommand{\desdevfy}{\frac{\de\,S}{\de\,\vfy} }
\newcommand{\dessdevfxdevfy}{\frac{{\de}^2\,S}{\de\,\vfx\,\de\,\vfy} }
\newcommand{\dessdevfxhdevfmxh}{\frac{{\de}^2\,S}{\de\,\vfxh\,\de\,\vfmxh} }

\newcommand{\mes}{< M \left|\left[j\,\left({\frac x 2}\right),
j\,\left(- {\frac x 2}\right)\right]\right| M > }
\newcommand{\mepm}{< M \left|j\,\left({\frac x 2}\right)
j\,\left(- {\frac x 2}\right)\right| M > }

\newcommand{\mepnr}{< M \left|j\,\left({\frac x 2}\right)
\right| p, n > }
\newcommand{\mepnl}{<  p, n \left|j\,\left(- {\frac x 2}\right)
\right| M > }
\newcommand{\mepp}{<  p^{'} \left|j\,\left(x\right)\right| p > }
\newcommand{\mepnrz}{< M \left|j\,\left(0\right)\right| p, n > }

\newcommand{\intzinf}{{\int\limits_{0}^{\infty}}}
\newcommand{\intinf}{{\int\limits_{-\infty}^{\infty}}}
\newcommand{\intminfz}{{\int\limits_{-\infty}^{0}}}
\newcommand{\intminfmm}{{\int\limits_{-\infty}^{- m}}}
\newcommand{\intxz}{{\int\limits_{- x_0}^{x_0}}}
\newcommand{\intm}{{\int\limits_{- m}^{m}}}

\newcommand{\intminf}{{\int\limits_{m}^{\infty}}}

\newcommand{\proon}{{\prod\limits_{1}^{n}}}
\newcommand{\proinj}{{\prod\limits_{i\neq j}}}

\newcommand{\cogb}{consideration given below }
\newcommand{\app}{approach }
\newcommand{\axc}{axiomatic }
\newcommand{\el}{elastic }
\newcommand{\es}{elastic scattering }
\newcommand{\am}{amplitude }

\newcommand{\sca}{scattering }
\newcommand{\sam}{scattering amplitude }

\newcommand{\esa}{elastic scattering amplitude }
\newcommand{\an}{analyticity }
\newcommand{\anl}{analytical }
\newcommand{\co}{commutative }
\newcommand{\nc}{noncommutative }
\newcommand{\ncy}{noncommutativity }
\newcommand{\coy}{commutativity }

\newcommand{\qftp}{quantum field theory.}

\newcommand{\con}{condition }

\newcommand{\loi}{Lorentz invariance }
\newcommand{\sams}{scattering amplitudes }
\newcommand{\fu}{function }
\def \op {operator \ }
\def \opp {operator. \ }

\newcommand{\pq}{p, q }
\newcommand{\pqpr}{p', q' }

\newcommand{\akrq}{a^{+}\,(q) }
\newcommand{\amqpr}{a^{-}\,(q') }

\newcommand{\ainqpr}{a_{in}\,(q') }
\newcommand{\ainkrq}{a^{+}_{in}\,(q) }
\newcommand{\ainkrqpr}{a^{+}_{in}\,(q') }

\newcommand{\aoutqpr}{a_{out}\,(q') }
\newcommand{\aoutkrq}{a^{+}_{out}\,(q) }

\newcommand{\ainkrqprz}{ a^{+}_{in}\,(q) a^{+}_{in}\,(p) | 0 > }
\newcommand{\ainkrqrp}{ a^{+}_{in}\,(q) | p > }

\newcommand{\deyzmxz}{{\delta}\,(\yz - \xz) }

\newcommand{\rqpin}{| q_{in}, p_{in} > }
\newcommand{\rp}{| p > }
\newcommand{\rpin}{| p_{in} > }
\newcommand{\rpout}{| p_{out} > }
\newcommand{\pqproutl}{< p'_{out}, q'_{out} | }
\newcommand{\pproutovaaoutqprl}{< p'_{out} |
a_{out}\,(q') }
\newcommand{\pqproutqpin}{< p'_{out}, q'_{out} |  q_{in}, p_{in} > }

\newcommand{\mepqprsrqp}{< p', q' | S | q, p > }

\newcommand{\mepqprakrqsrp}{< p', q' | a^+\,(q)\,S | p > }

\newcommand{\mepqprdesdevfxrp}{< p', q' | \desdevfx | p > }
\newcommand{\meppramqprdesdevfxrp}{< p' | a^-\,(q')\,\desdevfx | p > }
\newcommand{\mepprcamqprdesdevfxrp}{< p' | [a^-\,(q'), \desdevfx] | p > }
\newcommand{\mepqprcsakrqrp}{< p', q' | [S ,a^+\,(q)] | p > }

\newcommand{\mepqproutsqpout}{< p'_{out}, q'_{out} | S |
q_{out}, p_{out} > }
\newcommand{\mepqproutainkrqrp}{< p'_{out}, q'_{out} |
\ainkrq | p > }
\newcommand{\mepqproutaoutkrqrp}{< p'_{out}, q'_{out} |
\aoutkrq | p > }
\newcommand{\mepprjxaoutqprrq}{< p' | \jx\,\aoutqpr | q > }
\newcommand{\mepprpsxyrp}{< p' | \psxy | p > }
\newcommand{\mepprpsxmaymarp}{< p' | \psxmayma | p > }

\newcommand{\mepqproutpsxrp}{< p'_{out}, q'_{out} | \psx | p > }
\newcommand{\mepqproutpsoutxrp}{< p'_{out}, q'_{out} | \psoutx | p > }
\newcommand{\mepqproutjxrp}{< p'_{out}, q'_{out} | j\,(x) | p > }

\newcommand{\vfinx}{\varphi_{in}\,(x) }
\newcommand{\vfoutx}{\varphi_{out}\,(x) }
\newcommand{\dvfinx}{\dot{\varphi}_{in}\,(x) }

\newcommand{\fsqx}{f_{q}\,(x)}
\newcommand{\fsqxi}{f_{q}\,(\xit)}
\newcommand{\fsqxn}{f_{q}\,(\xn)}

\newcommand{\fsqsxh}{f^{*}_{q}\,(\frac x 2)}
\newcommand{\fsqprx}{f_{q'}\,(x)}
\newcommand{\fsqprsx}{f^{*}_{q'}\,(x)}
\newcommand{\fsqprsxh}{f^{*}_{q'}\,(\frac x 2)}

\newcommand{\fsnxoxn}{f_{n}\,(\xo, \cdots \xn)}
\newcommand{\fsnxoxixn}{f_{n}\,(\xo, \cdots \xit = x,  \cdots \xn)}

\newcommand{\dtqp}{d^3\,q' }

\newcommand{\exmiqx}{e^{- i\,q\,x}}

\newcommand{\exipprmpa}{e^{ i\,(p' - p)\,a}}
\newcommand{\exiqpqprxh}{e^{ \frac{i\,(q + q')\,x}{2}}}

\newcommand \znam {{(2\,\pi)}^{\frac 3 2} \,\sqrt{2\,\qz} }

\newcommand{\dfsqx}{\dot{f}_{q}\,(x)}
\newcommand{\ddfsqx}{\ddot{f}_{q}\,(x)}

\newcommand{\dtx}{d^3\,x }
\newcommand{\dfxo}{d^4\,x_{1} }
\newcommand{\dfxi}{d^4\,x_{i} }
\newcommand{\dfxn}{d^4\,x_{n} }

\newcommand{\al}{\alpha }
\newcommand{\ral}{| \al > }

\newcommand{\bet}{\beta }

\newcommand{\dvfx}{\dot{\varphi}\,(x) }

\newcommand{\psx}{{\psi}\,(x)}
\newcommand{\psxy}{{\psi}\,(x, y)}
\newcommand{\psxmayma}{{\psi}\,(x - a, y - a)}
\newcommand{\psoutx}{{\psi}_{out}\,(x)}

\newcommand{\ppr}{p' }
\newcommand{\qpr}{q' }

\newcommand{\ddvfx}{\ddot{\varphi}\,(x) }
\newcommand{\dfsqprsy}{\dot{f}^{*}_{q'}\,(y)}
\newcommand{\fsqprsy}{f^{*}_{q'}\,(y)}
\newcommand{\dvfy}{\dot{\varphi}\,(y) }
\newcommand{\dty}{d^3\,y }
\newcommand{\dfy}{d^4\,y }

\newcommand{\sumzn}{{\sum\limits_{n = 0}^{n}}}
\newcommand{\sumzi}{{\sum\limits_{n = 0}^{\infty}}}

\newcommand{\vfxo}{\varphi\,(x_1) }

\newcommand{\vfxn}{\varphi\,(x_n) }
\newcommand{\vfxnmo}{\varphi\,(x_{n - 1}) }
\newcommand{\vfxi}{\varphi\,(x_i) }

\newcommand{\vfxoxn}{\vfxo \cdots \vfxn }
\newcommand{\xn}{x_n }
\newcommand{\xit}{x_i }
\newcommand{\tod}{\theta_{12}}

\newcommand{\be}{\begin{equation}}
\newcommand{\ee}{\end{equation}}
\newcommand{\bn}{\begin{eqnarray}}
\newcommand{\en}{\end{eqnarray}}
\newcommand{\bnn}{\begin{eqnarray*}}
\newcommand{\enn}{\end{eqnarray*}}
\newcommand{\ba}{\begin{array}}
\newcommand{\ea}{\end{array}}

\begin{document}
\begin{flushright}
HIP-2003-37/TH
\end{flushright}

\begin{center}

{\Large{\bf{Analyticity and Forward Dispersion Relations in Noncommutative
Quantum Field Theory}}}
\vskip .7cm

{\bf{\large{M. Chaichian$^a$, M. N. Mnatsakanova$^{\dagger}$,
A. Tureanu$^a$ \ \ and \ \ Yu. S. Vernov$^{\dagger\dagger}$}}}

{\it $^a$High Energy Physics Division, Department of
Physical Sciences,
University of Helsinki\\
\ \ {and}\\
\ \ Helsinki Institute of Physics,
P.O. Box 64, FIN-00014 Helsinki, Finland\\

$^{\dagger}$Institute of Nuclear Physics, Moscow State University,\\
Moscow, 119899 Russia\\

$^{\dagger\dagger}$Institute for Nuclear Research,
Russian Academy of Sciences,\\Moscow, 117312 Russia }

\setcounter{footnote}{0}

{\bf Abstract}

\end{center}

We derive the analytical properties of the elastic forward scattering amplitude of two scalar particles from
the axioms of the noncommutative quantum field theory. For the case of only space-space noncommutativity, i.e.
$\theta_{0i}=0$, we prove the dispersion relation which is similar to the one in commutative quantum field
theory. The proof in this case is based on the existence of the analog of the usual microcausality condition
and uses the Lehmann-Symanzik-Zimmermann (LSZ) or equivalently the Bogoliubov-Medvedev-Polivanov (BMP) reduction
formalisms. The existence of the latter formalisms is also shown. We remark on the general noncommutative
case, $\theta_{0i}\neq0$, as well as on the nonforward scattering amplitude and mention their peculiarities.

\section  {Introduction}

The proof of the \anl properties of \sams is one of the most remarkable
achievements of the \axc \app to quantum field theory. The dispersion relations (DR) for the \esa
were derived in the works of Gell-Mann, Goldberger, Thirring,
Miyazawa, Nambu and Oehme \cite{1}-\cite{4}. They were rigorously proven in the works of
Bogoliubov \cite{5}, Oehme \cite{6}, Symanzik \cite{7}, Bremermann, Oehme, Taylor \cite{8} and
Lehmann \cite{9}. The detailed proof of DR was given in the book of Bogoliubov,
Medvedev and Polivanov \cite{10}.

The implications of the modern ideas of noncommutative geometry \cite{Connes} in physics have been lately of
great interest, though attempts can be traced back as far as 1947 \cite{Snyder}. Plausible new arguments for studying  \nc quantum field theories (NC QFT) \cite{SW, Shahin, Dopli} (for a review,
see \cite{DN}) render
the problem of analyticity in such theories actual. However, the task of establishing the analytical
properties of \nc field theory is highly nontrivial. In passing from a usual space-time
manifold to a space on which the coordinate operators do not commute, i.e.
\be
[x_\mu,x_\nu]=i\theta_{\mu\nu},
\ee
where $\theta_{\mu\nu}$ is an antisymmetric constant matrix of dimension (length)$^2$,
the interactions acquire a nonlocal character and at the same time the Lorentz invariance is lost. It is mainly this nonlocal
nature which gives rise to a novel behaviour of the NC QFT. For the derivation of dispersion relations, of
crucial importance is the microcausality, which is affected by the \ncy of space-time. The effect is drastic
when time does not commute with the spacial coordinates ($\theta_{0i}\neq0$), in the sense that
microcausality is completely lost \cite{LAG, 17} (see also \cite{SST} for acausal macroscopic
effects in scattering). In the
case of theories with commutative time ($\theta_{0i}=0$) microcausality survives, but as a weaker condition
than in the commutative case \cite{LAG} (see eq. (\ref{2})). For this reason one may hope that dispersion
relations can still be obtained in field theories with only space-space noncommutativity.

The first step in this direction was made by Liao and Sibold
\cite{11}. The essential difference between the analytical properties
of the scattering amplitude in commutative and noncommutative
cases found in their work was related to a specific way of continuation
of the scattering amplitude to the complex plane. As a result, in \cite{11} it was concluded that a
derivation of the DR was not possible.

In the present work we aim at deriving DR first for forward elastic scattering of two spinless particles
with masses $m$ and $M$. In the case of scattering of particles with spin, such as $\pi N$-scattering,
our considerations refer to the invariant amplitude of those processes. We prove that
if the \ncy affects only the space variables\footnote{The same case of \ncy was considered in \cite{11}.},
i.e. when $\tezi = 0$, then the standard DR with $n$ subtractions, analogous to the commutative case,
are valid.

In the case of space-space noncommutativity we can choose the coordinates in such a way
that only $\teod = - \tedo \neq 0$.
Then the usual condition of local commutativity can be substituted by
its analog containing only the $\xz$ and $\xt$ coordinates \cite{LAG} (see eqs. (\ref{1})
and (\ref{2})).

We show that the above-mentioned noncommutative analog of local commutativity
is indeed sufficient for proving the same analytical properties of the forward \esa as in the commutative case.
We admit that, similarly to the \co case, the \sam is bounded by a polynomial (\ref{5}).
However our proof is valid under a weaker \con (\ref{23_1}) than is usually used.
Specifically, we substitute the \con of polynomial boundedness on the
\sam by anything less than an exponential growth.

%
%
In the general case ($\tezi \neq 0$), the analyticity issue is rather obscure due to
the lack of \nc analog of local commutativity. Besides, the existence of reduction
formulas, which is the basis for the proof of analyticity, is not clear. Nevertheless,
if reduction formulas survive in this case, we come to the conclusion that
in the relations which follow from analyticity, the appearance of an additional term
is very likely.

We have proven the \an of the \el \sca amplitudes on the basis of
Lehmann-Symanzik-Zimmermann (LSZ) reduction formulas \cite{LSZ}. In the end of the
paper we show that the same results can alternatively be derived using the
Bogoliubov-Medvedev-Polivanov (BMP) approach \cite{10}.

In the Appendix the status of the reduction formulas in NC space-space theory
is considered.

\section  {Forward scattering}

We shall study the problem of \an of forward \es \am  in case of \nc \qftp

We consider the case when time commutes with the space variables, $\theta_{0i}=0$, and
restrict ourselves to the \sca of two scalar particles with masses $m$ and
$M$.

In the commutative case we admit the \con of local commutativity:
\be\label{1}
\kjxy = 0, \quad \mbox{if} \quad \xmys < 0,
\ee
where $\jx$  is the current of interacting fields.

The local \coy \con (\ref{1}) is an independent axiom  and not the consequence of Lorentz
invariance. This \con means the absence of infinite speed of any
interaction propagation. \loi gives us the possibility to write this \con
in an invariant form.

In the \nc case with $\theta_{0i}=0$, the Lorentz symmetry $SO(3,1)$ is broken to $SO(1,1)\times SO(2)$
\cite{LAG} and we can
choose the coordinates in such a way that only $\teod = - \tedo \neq 0$. In the
direction perpendicular to the \nc $(x_1,x_2)$-plane we admit the existence of the
maximal speed of interactions propagation. Then on the same basis as in
the usual case, we assume the local commutativity condition (or microcausality) to be \cite{LAG}
\be\label{2}
\kjxy = 0 \quad \mbox{if} \quad  \xzmyzs - \xtmyts < 0.
\ee
This condition was shown to be valid \cite{17} for $x_0=y_0$ using the equal-time commutation relations
for the cases when $j(x)$ is any power of field operators with $\star$-product. Due to the remaining SO(1,1)
symmetry, this implies the validity of (\ref{2}) in the whole region $\xzmyzs - \xtmyts < 0$.

\subsection{Analyticity in the framework of LSZ approach}

If in the \nc case "in" and "out" fields can be constructed in the same way as in
usual theory then the standard Lehmann-Symanzik-Zimmer\-mann (LSZ)
reduction formulas are valid (see the Appendix) and the \sam is:
\be\label{3}
\fevq = \int\,d^4x\,\exs\,\taxz\,\fx,
\ee
where
$$
\fx = \mes, \qquad j\,(x) \equiv (\Box + m^2)\,\vf\,(x).
$$
We omit in (\ref{3}) numerical factors which are irrelevant to the analytical
properties of $\fevq$. Eq. (\ref{3}) is written in the reference frame in which the
particle with the mass $M$ is at rest.  $E$ and $\vq$ are the energy and
momentum of the particle with mass $m$.

Actually $\fx$ contains an additional term:
$$
\delta\,(x_0)\,< M \left|\left[j\,\left({\frac x 2}\right),
{\frac {\partial} {\partial x_0}}\,\vfmxh \right]\right| M >,
$$
which does not change the analytical properties of $\fevq$.
The contribution of this term in (\ref{3}) is some polynomial in $E$. To
show this it is sufficient to admit the standard assumption that $\jx$
is some polynomial of $\vfx$ and use equal time commutation relations
(see e.g. \cite{8}, eq. (2.2) or \cite{12}, chapter 18).

Precisely speaking, the matrix element in (4) is an operator-valued
generalized function (see, e.g. \cite{23ad} and [10]). The corresponding
questions are not specific to the NC case and that is why we do not dwell
on them any further. We only mention that as our proof is valid under a
weaker condition than polynomial boundedness of the scattering amplitude
(see the condition (31)), we can consider the class of generalized functions to be 
more general than the tempered distributions.

In order to extend $\fevq$ to the upper complex $E$-plane ($Im\,E > 0$) we integrate (\ref{3})
over $x_1$ and $x_2$ (similarly as in \cite{11}).

Then, using (\ref{2}), $\fevq$ is represented in the form:
\be\label{4}
F\,(E, |{\vec{q}}|, \vec{e}) = \intzinf \exiexz\,d\,x_0\,\intxz
\exxt\,\Phi\,(x_0, x_3)\,d\,x_3,
\ee
where
$$
\Phi\,(x_0, x_3) = \int\,\fx\,\exq\,d\,x_1\,d\,x_2.
$$
As shown in \cite{LAG} in space-space \nc theory, due to the $SO(1,1)$ symmetry, $q_0^2 - q_3^2 = const$.
In the usual (commutative) case one has from the energy-momentum relation
$$
E_0^2 \equiv E^2-q_3^2= m^2 + q_1^2 + q_2^2.
$$
In the \nc case the energy-momentum relation is altered. However, the explicit
expression for $E_0^2$ is not essential for our analyticity considerations.

In order to exclude the singularity at $\sqez$, we make the substitution
$$
F\,(E,|{\vec{q}}|, \vec{e}) \rightarrow  \frac 1 2 \,\left(F\,(E, |{\vec{q}}|,
\vec{e}) + F\,(E, |{\vec{q}}|, - \vec{e}\right) \equiv \fe.
$$
(This is a standard procedure, see \cite{13}, chapter 10.)
In accordance with (\ref{4}) and writing $\vec{q}$ in the form $\vec{q} = \vec{e}\,|{\vec{q}}|, \;
|{\vec{e}}| = 1$, we obtain
\be\label{4'}
\fe = \intzinf\,\exiexz\,d\,\xz\,\intxz\,\cosxt\,\Phi\,(x_0,
x_3)\,d\,x_3.
\ee
A direct extension of $\fe$ into the complex $E$-plane is impossible since
$$
Im\,\sqez > Im\,E
$$
(see \cite{13}, chapter 10).

To overcome this obstacle, following \cite{13}, we substitute $\fe$ by the
regularized amplitude $\fee$:
\be\label{4''}
\fee =
\intzinf\,\exiexz\,d\,\xz\,\intxz\,\cosxt\,e^{-
\varepsilon\,(x_0^2 + x_3^2)}\,\Phi\,(x_0, x_3)\,d\,x_3.
\ee
$\fee$ is an analytical function in the upper half-plane, where the integral in
(\ref{4''}) converges.

The main problem is to prove the existence of the analytical function
$\fe = \lim\limits_{\varepsilon \rightarrow 0}\,\fee$. To this end we shall use the
analytical properties of $\fee$. Our goal is to represent  $\fee$ in the
complex $E$-plane as an integral over the real axis only and then take the limit $\varepsilon\rightarrow
0$. But it is impossible  to do this directly as $\fee \not \to 0$ as $E
\to \infty$. So first we have to construct a function which would have this property.
We admit that there exists a number $n$ such that
\be\label{5}
\frac{\fe}{E^n} \to 0 \quad  \mbox{as}   \quad    E \to +\infty.
\ee
In the commutative case one takes $n = 2$ in accordance with the Froissart-Martin bound \cite{Froissart,
Martin, Martin2}, but
here we have to admit a more general condition. Evidently, from (\ref{5}) it follows that
$$
\frac{\fee}{E^n} \to 0 \quad  \mbox{as}   \quad    E \to +\infty.
$$

Condition (\ref{5}) is valid also as $E
\to - \infty$ since
\be\label{6}
F\,(- E + i\,0) = F^{*}\,(E + i\,0), \quad
F_{\eps}\,(- E + i\,0) = F_{\eps}^{*}\,(E + i\,0).
\ee
Eq. (\ref{6}) is the standard crossing symmetry condition.  We point out that
$j\,(x)$ is a Hermitian operator.

Evidently, the function
$$
\psee = \frac{\fee}{\proon\,\emei},    \qquad E_i > E_0
$$
satisfies the condition
\be\label{7}
\psee \to 0, \qquad E \to\pm\,\infty.
\ee
Thus we can use the Cauchy formula,
\be\label{8}
\psee = \frac 1 {2\,\pi\,i}\,\int_C\,\frac{\pseep\,d\,E'}{E' - E},
\quad Im\,E > 0,
\ee
where $C$ consists of the interval $(- R, R)$, excluding $n$ arbitrarily small
semicircles around $E_i$, and a semicircle in the upper half-plane.

Now we shall demonstrate that, due to the local commutativity condition (\ref{2}),
\be\label{7'}
\pser \to 0, \quad \mbox{if} \quad R \to \infty, \quad 0 < \varphi < \pi.
\ee
Indeed, if $|E| \to \infty$, then
\be\label{9}
Im\,\sqez \cong Im\,E - Im\,{\frac{E_0^2}{2\,E}}.
\ee
Thus
\be\label{10}
\left| {\exiexz}\,\cosxt\right| \leq e^{- Im\,E\,(\xz - \xt)}
e^{\frac{E_0^2\,|\xt|\,\sin\,\vf}{R}}.
\ee
The first factor on the r.h.s. of (\ref{10}) is less than
unity as $\xz > \xt$. Due to the factor $\exp(- \eps\,{\xt}^2)$, in (\ref{4''}) the integral over $\xt$ converges
when $\xz \to \infty$, so the integration is actually over some
finite interval $(- \bar{x}_3(\epsilon), \bar{x}_3(\epsilon))$. Thus the second factor tends to unity at any fixed $\eps$
if $R \to \infty$. Thus the growing factor in the integrand in eq.  (\ref{4''}) disappears as $|E| \to \infty$.
We can thus conclude that condition (\ref{7'}) follows from the condition (\ref{7}).

Actually in order to prove that condition (\ref{7'}) is the consequence of
condition (\ref{7}), it is sufficient to assume that $\pser$ grows more slowly than
any exponent and use the Phragmen-Lindel\"of theorem (see e.g. \cite{14}). This
is a very weak requirement on the behaviour of the scattering amplitude at infinite energies.
Any function, which grows even as
$\exp\,(R^{\alpha}), \; 0 < \alpha < 1$, satisfies it.

Thus we can put $R = \infty$ in (\ref{8}). So
\be\label{11}
\psee = \frac 1 {2\,\pi\,i}\,\intinf\,\frac{\pseep\,d\,E^{'}}{E^{'} -
E} - {\frac 1 2}\,\sum\limits_{i =
1}^{n}\,\frac{\feei}{\eime\,\proinj\,\eimej}, \quad Im\,E > 0.
\ee
Eq. (\ref{11}) is valid at any fixed $\eps$. Now we shall take the limit $\eps \rightarrow 0$.
First we consider the interval $(m, \infty)$. If $\ep > \ez$, we can go
to the limit  $\eps \rightarrow 0$ without any problem as in this interval
$\lim\limits_{\eps \to 0}\,\feepr = \fepr$ (see eqs. (\ref{4'}) and (\ref{4''})). In the
interval $(m, \ez)$ we can not use (\ref{4''}). But this interval is a
physical one and so (\ref{3}) and (\ref{4}) coincide in this interval.

Thus
$$
\feepr = \int\,\exieprxz\,\cos{(\vec{q}\vec{x})}\,\exp(-
\eps\,({\xz}^2 + {\xt}^2))\fx\,\dfx,
$$
and we see that $\feepr \to \fepr$ as $\eps \to 0$.

We stress that only this interval is specific for the NC case. The interval
$(- \infty, - m)$ can be treated similarly in accordance with (\ref{6}).

The remaining interval can be considered as in the commutative case (see e.g.
\cite{13}).

To handle this interval, we shall construct the analytical function in the
lower half-plane and then prove that this function is an analytical
continuation of $\fee$. To this end we use the function
\be\label{13}
\tfevq =  \int\,d^4x\,\exs\,\tamxz\,\fxm.
\ee
Then we substitute $\tfevq$ by
\be\label{13'}
\tfee = \int\,d^4x\,\cosqx\,\tamxz\,e^{- \eps\,(x_0^2 + x_3^2)}\,\fxm.
\ee
Evidently
\be\label{14}
\tfeemiz = \femepiz = \fesepiz.
\ee
The last equality in (\ref{14}) is eq. (\ref{6}). To prove the first equality it is
sufficient to replace $x$ by $- x$. In (\ref{14}) we can put $\eps = 0$ and obtain
\be\label{14'}
\tfemiz = \fsepiz.
\ee
Similar to eq. (\ref{4''}) we have
\be\label{15}
\tfee =
\intminfz\,\exiexz\,d\,\xz\,\intxz\,\cosxt\,e^{-
\varepsilon\,(x_0^2 + x_3^2)}\,\tilde \Phi\,(x_0, x_3)\,d\,x_3.
\ee
The function
$$
\tpsee = \frac{\tfee}{\proon\,\emei},     \qquad E_i > E_0
$$
is an \anl  function in the lower half-plane and $\tpsee \to 0, \; E
\to\pm\,\infty$. We use the same arguments as for the proof of \an of $\psee$
in the upper half-plane. Thus
\be\label{16}
\frac 1 {2\,\pi\,i}\,\int_{\tilde C}\,\frac{\tpseep\,d\,E^{'}}{E^{'} - E}
= 0,     \quad Im\,E > 0,
\ee
where $\tilde C$ consists of the interval $(R, - R)$, excluding $n$ arbitrarily small
semicircles around $E_i$ and a semicircle in the lower half-plane.

We shall now sum up the
expressions (\ref{8}) and (\ref{16}). Using (\ref{14}) and taking into account that the
integral over a semicircle in the lower half-plane tends to zero if $R \to
\infty$ for the same reason as the corresponding integral in the upper
half-plane, we obtain that
\bn\label{17}
\psee = \frac 1 {\pi}\,\intminf\,\frac{Im\,\pseep\,d\,E^{'}}{E^{'} -
E} + \frac 1 {\pi}\,\intminfmm\,\frac{Im\,\pseep\,d\,E^{'}}{E^{'} - E} \cr
-\sum\limits_{i = 1}^{n}\,\frac{Re\,\feei}{\eime\,\proinj\,\eimej} +
\frac 1 {2\,\pi\,i}\,\intm\,\frac{\left(\pseep - \tpseep\right)\,d\,E^{'}}
{E^{'} - E}\ ,\,
 Im\,E > 0.
\en
In the first three terms in (\ref{17}) we can go to the limit $\eps \rightarrow 0$. In order to be able to
take the corresponding limit in
the remaining integral, we shall first obtain in the physical domain the expression
for $\fevq - \tfevq$, suitable to extension for nonphysical $E$, i.e. $- m < E < m$.
>From the definitions (\ref{3}) and (\ref{13}) it follows that
\be\label{18}
\fevq -\tfevq = \fpevq - \fmevq,
\ee
where
\be\label{19}
\fpmevq =  \int\,d^4x\,\exs\,\fpmx,
\ee
\be\label{20}
\fpx = \mepm, \quad      \fmx = \fpxm.
\ee
Assuming that the vectors $|p, n>$ form a complete set of basis
vectors, we have
\be\label{21}
\mepm
=\sum\limits_n\,\sum\limits_{p^0_n}\,\int\,d^3\,p\,\mepnr\,\mepnl,
\ee
where $p$ stands for the momentum of the state, $p^0_n$ is the energy of the state $|p, n>$ and $n$
denotes all other quantum numbers.

Using the equality
\be\label{22}
\mepp = \expa\,< p^{'} \left|j\,\left(x - a\right)\,\right| p >,
\ee
where $|\,p\,>$ and $|\,p^{'}\,>$ are eigenvectors of
the operator $p$, we see that,
due to (\ref{20}) and (\ref{21})
\be\label{23}
\fpmevq =  \sum\limits_n\,\sum\limits_{p^0_n}\,{\left|\mepnrz\right|}^2
\,\delta\,(p^0_n - M \mp E), \quad  \vec{p} = \mp \vec{q}.
\ee
Thus $\fpmevq \neq 0$ only if
\be\label{24}
\sqrt{M^2_n + {\vec{q}}^2} = M \pm E, \quad  p^0_n = \sqrt{M^2_n +
{\vec{q}}^2}.
\ee
Let us assume that (as e.g. in the case of $\pi N$-scattering) $M_n \geq M + m$,
thus excluding the one-particle intermediate state, $M$. We can extend the expression
(\ref{23}) for $E$ in the interval $(- m, m)$. The functions $\fpmevq \neq 0$ in
this interval if
$$
\sqrt{M^2_n + E^2 - m^2} = M \pm E,
$$
which is possible only if $M_n = M$ and $E = - {m^2} / {2\,M}$.

Thus we see that in the integral under consideration (excluding two points: $\pm\,\frac{m^2}{2\,M}$)
$$
\lim\limits_{\eps \rightarrow 0}\,(\psee - \tpsee) = 0,
$$
as
$$
\fpevq - \fmevq = 0.
$$
In order to make the integral over the interval $(- m, m)$ vanish, it is sufficient
to substitute $\psee$ and $\tpsee$ by the functions:
$$
\bfee = \frac{E^2 - \frac{m^4}{4\,M^2}}{(E - E_{n + 1})(E - E_{n + 2})}\,
\psee, \quad  E_{n + 1} > E_0, \; E_{n + 2} > E_0,
$$
$$
\tbfee = \frac{E^2 - \frac{m^4}{4\,M^2}}{(E - E_{n + 1})(E - E_{n + 2})}\,
\tpsee, \quad  E_{n + 1} > E_0, \; E_{n + 2} > E_0.
$$

Representing $\bfee$ by an expression analogous to (\ref{17}), we see that there
exists $\lim\limits_{\eps \rightarrow 0}\,\bfee = \bfe$. Moreover $\tbfe =
\lim\limits_{\eps \rightarrow 0}\,\tbfee$ is an \anl continuation of $\bfe$.
The function $\bfe$ and consequently the function $\left(E^2 - \frac{m^4}{4\,M^2}\right)\fe$
are \anl in the whole $E$-plane excluding the cuts $(- \infty, -
m), \: (m, \infty)$. $\fe$ is an \anl function in the same domain
excluding the points $\pm\,\frac{m^2}{2\,M}$, where it has poles.

Finally, using (\ref{6}) and (\ref{14'}), we arrive at the usual expression for
$\fe$:
\be\label{25}
\fe = \frac
{2\,E^n}{\pi}\,\intminf\,\frac{Im\,\fepr\,d\,E^{'}}{{(E^{'})}^{n -
1}({E^{'}}^2 - E^2)} + \sum \limits _{\stackrel {k = 0,}{even}}^{n -
2}\,C_k\,E^k + \mbox{pole terms}, \quad Im\,E \neq 0.
\ee
In the limit $Im\,E \rightarrow 0$, (\ref{25}) becomes the usual dispersion relation.

We can conclude that if the LSZ reduction formulas are valid in NC field
theory and the condition of local commutativity can be replaced by the
condition (\ref{2}), the NC forward scattering amplitude has the same analytical
properties as in the commutative case.

We would like to point out that our proof of analyticity of the forward scattering amplitude presented above
remains still valid if one allows asymptotically a growth of the amplitude $F(E)$ much faster than a
polynomially bounded one. Indeed, it is sufficient to assume that there exists $\al, \; 0 < \al <1$
such that
\be\label{23_1}
\left|\,\fe \,\right| < \exp\,(E^{\al}), \qquad E \to \infty.
\ee
Consequently, the function
\be\label{24_1}
\psee = \fee\,\kse,
\ee
where
$$
\kse = \exp\,\left[- \left({\sqrt{m^2 - E^2}}\right)^{\beta}
\exp\left(- i\,\pi\,\beta\right) \right], \; 0 < \beta < 1, \; \al < \beta,
$$
satisfies the necessary condition (\ref{7}).

In fact, for any $\vf, \: 0 < \vf < \pi$, we have
$$
\left|\kse\right| < \exp\,\left[- {|E|}^{\beta}
\cos\left(\vf - \frac{\pi}{2}\right)\,{\beta}\right].
$$
Then it is easy to see that $\kse$ is an \anl \fu in the whole $E$-plane with
cuts $(m, \infty), \; (- \infty, - m)$ satisfying the conditions
$$
\xi\,(- E + i\,0) = \xi^{*}\,(E + i\,0) = \xi\,(E - i\,0).
$$
One can check that all the previous steps in the proof go through also for the new
function $\psee$.

\subsection{Analyticity in the framework of BMP approach}

The same results can be obtained on the basis of Bogoliubov-Medvedev-Polivanov
(BMP) \cite{10} reduction formulas and by using the analog of Bogoliubov microcausality
condition \cite{10,13} in
the NC case. For the forward scattering, the reduction formula is:
\be\label{26}
\fevq = \int\,d^4x\,\exs\,\fxc,
\ee
where
\be\label{27}
\fxc = < M \left| \dessdevfxhdevfmxh\,S^* \right| M >.
\ee
We replace now the role of the local commutativity condition (\ref{2}) by the {\it modified}
Bogoliubov microcausality condition:
\be\label{28}
\dedevfx \,\jy = 0, \quad \mbox{if} \quad \xz <\yz  \quad \mbox{or} \quad
\xzmyzs - \xtmyts < 0,
\ee
where
$$
\jx \equiv i\,\desdevfx\,S^*, \quad \jx = \jsx
$$
is the current in the BMP \axc
approach.  The condition $\xz < \yz$ coincides with the corresponding original
condition in the commutative case. The condition $\xzmyzs - \xtmyts < 0$ substitutes the
usual ${(x - y)}^2 < 0$.

To extend $\fevq$ in the upper and lower half-planes, we use
\be\label{29}
\fraevq = \int\,\dfx\,\exs\,\frax
\ee
correspondingly, where
$$
\frx = < M \left| \dedevfmxh\,\left(\desdevfxh\,S^*\right) \right| M >,
\quad \fax = \frmx.
$$
In accordance with (\ref{28}), $\frx = 0$ if $\xz < 0$ or $x_0^2 < x_3^2$.
In this formalism the proof of the analytical properties of $\fevq$ is the same
as in LSZ formalism. We only need to show that $\fevq - \frevq = 0$ at
physical energies. This can be done in the same way as (\ref{23}) has
been obtained. It is easy to show that
$$
\fevq - \frevq \sim \de\,\left(\sqrt{M^2_n + {\vec{q}}^2} + E - M\right) =
0,  \quad \mbox{for} \quad M_n \geq M.
$$
We point out that at physical energies the equality of the usual and
retarded amplitudes is valid for nonforward scattering as well (see \cite{13}).

\section  {Comments on the general NC case $\theta_{0i}\neq 0$}

In the following we shall briefly consider the general case $\tezi \neq 0$. Let us
assume (though this assumption may be not valid) that the LSZ reduction formula (\ref{3}) is valid also in this
case, but we do not have the local commutativity condition (\ref{2}). In \cite{17} it was shown that
condition (\ref{2}) is indeed not fulfilled in this case.

The function $\fee$ is an \anl function as before, but, even if the polynomial boundedness condition
(\ref{5}) is valid, we can not exclude the possibility that this function grows
exponentially in the whole complex plane.  If (\ref{5}) exists then we can obtain DR
for forward direction, but with
an additional term. Indeed, DR follows from a similar relation at finite
$R$ by taking the limit $R \rightarrow \infty$. Now we can also put $R = \infty$,
but we have no arguments that the integral in the complex plane tends to zero
if $R \to \infty$.  Nevertheless this integral converges as the integral
over the real axis converges and $E$ is the physical energy.

We point out that we have to work with a regularized amplitude and only
at the end go to the $\eps \rightarrow 0$ limit. Although in the \co case the \anl properties of the \sam
in the absence of local commutativity conditions have been studied
\cite{18, 19}, in the NC case with $\theta_{0i}\neq0$ the issue requires further investigation.

\section{Conclusions}

We have derived the analytical properties for the forward elastic scattering amplitude of two spinless particles on a
space-time with space-space noncommutativity ($\theta_{0i}=0$). Based on the
axioms of \nc quantum field theory and using as an essential ingredient a microcausality (local commutativity) condition
analogous, but weaker than the one in the commutative case, we have
proven the existence of forward dispersion relations for the above-mentioned type of noncommutativity. The
proof has been given using the Lehmann-Symanzik-Zimmermann (LSZ) formalism and it has been also shown to
hold by using the
Bogoliubov-Medvedev-Polivanov (BMP) formalism. In  both frameworks, the existence of the reduction formulas for the
space-space noncommutativity has also been demonstrated.

For the general noncommutative case, $\theta_{0i}\neq0$, however, the microcausality (local
commutativity) condition does not exist anymore and the existence of the reduction formulas is also doubtful.

As for the case of the elastic nonforward scattering amplitude, the number of independent kinematical
variables increases, becoming five for the general noncommutativity case $\theta_{0i}\neq0$ \cite{CMT}, as
compared with two in the usual commutative case. Thus the analytical properties of the scattering amplitude in
the nonforward case require a special investigation.

Finally, we would like to recall that in the commutative case, the existence of dispersion relations is the key step in
extending the analyticity domain from the Lehmann ellipse \cite{9} to the enlarged Lehmann-Martin
ellipse \cite{Martin2} and in ultimately deriving the high-energy Froissart-Martin
bound \cite{Froissart, Martin} on the scattering amplitude from the axioms of quantum field theory.
The aim for similar applications lies behind the idea of the present paper
(see also \cite{CMT}), which can be regarded as the first step along that direction. We hope to present
further results in a future communication.
\vskip .5cm
{\Large\bf Acknowledgements}

We are grateful to Andr\'e Martin, Claus Montonen and Vladimir Petrov for
discussions and useful comments.

\appendix
\section{Appendix: LSZ and BMP reduction formulas for NC quantum field theory}
\subsection {LSZ (Lehmann-Symanzik-Zimmermann) reduction formulas}
In the following we shall demonstrate that all the steps in deriving the LSZ reduction formulas for the
space-space case of noncommutativity are the same as in the commutative case (\cite{LSZ}, see also
\cite{12}). To this end, we shall
study \esa of two scalar particles
\be\label{1_2}
\pqproutqpin = \mepqproutsqpout.
\ee
However, the \cogb is quite general.

In order to exclude contribution of the unit \op in
(\ref{1_2}) ($S = 1 + iT$), we shall put $\pqpr \neq \pq$ .

Evidently
$$
\rqpin = \ainkrqprz = \ainkrqrp.
$$
Under the stability of one-particle, we have states $\rpin = \rpout = \rp$.
We represent $\ainkrq$ as integral of $\vfinx$ and $\dvfinx$. To this
end we use the mode decomposition of $\vfinx$
\bn\label{2_2}
\vfinx &=& \int\,\left(\ainkrqpr\,\fsqprsx + \ainqpr\,\fsqprx\right)\,\dtqp,\cr
\fsqx &=& \frac {\exmiqx}{\znam},
\en
where $\ainqpr$ is the annihilation \opp From (\ref{2_2}) it follows that
\be\label{3_2}
\dvfinx = i \,\int\,\qzpr\,\left(\ainkrqpr\,\fsqprsx -
\ainqpr\,\fsqprx\right)\,\dtqp.
\ee
Multiplying the combination $i\,\qz\,\vfinx + \dvfinx$ by $\fsqx$,
integrating this expression over $x$ and using (\ref{2_2}) and (\ref{3_2}), we obtain
that
\be\label{4_2}
\ainkrq = i \,\int\,\left(\vfinx\,\dfsqx - \dvfinx\,\fsqx\right)\,\dtx.
\ee
We substitute $\vfinx$ by $\vfx$, where$\vfx$ is an interacting field, so that
for any states $\ral$ and $|\beta>$
\be\label{5_2}
<\al | \vfx - \vfinx | \bet> \to 0 \quad \mbox{at} \quad \xz \to - \infty.
\ee

Thus
\be\label{4'_2}
\ainkrq = i \,\int\limits_{\xz \to  - \infty}\,\left(\vfx\,\dfsqx
- \dvfx\,\fsqx\right)\,\dtx.
\ee


Now we shall use the general equality
\be\label{6_2}
\int\, [ \; ]|_{\xz = - \infty}\,\dtx =
\int\, [ \; ]|_{\xz = + \infty}\,\dtx -
\int\,\frac{\partial}{\partial\,\xz}\,[ \; ]\,\dfx
\ee
with the integrand
\be\label{7_2}
[ \; ] = \mepqproutainkrqrp = i \,\int\,\mepqproutpsxrp|_{\xz = -
\infty}\,\dtx,
\ee
where $\psx \equiv \vfx\,\dfsqx - \dvfx\,\fsqx$.

First we show that the first term in the r.h.s. of (\ref{6_2}) is zero. Indeed,
$$
i \,\,\mepqproutpsxrp|\,\int_{\xz = + \infty},\dtx =
$$
$$
i \,\int\,\mepqproutpsoutxrp|_{\xz = + \infty}\,\dtx =
\mepqproutaoutkrqrp,
$$
where we have used (\ref{4_2}) for the out-field. But
$$
\mepqproutaoutkrqrp = 0,
$$
since $q \neq \ppr, \qpr$.
For the remaining term, by direct calculations we obtain
$$
\frac{\partial}{\partial\,\xz}\,\left[\,\vfx\,\dfsqx -
\dvfx\,\fsqx\,\right] = \vfx\,\ddfsqx - \ddvfx\,\fsqx =
$$
$$
\vfx\,\left({\nabla}^2 - m^2\right) \,\fsqx - \ddvfx\,\fsqx \quad
\mbox{as} \quad \left({\Box} + m^2\right) \,\fsqx = 0.
$$
Using integration by parts we have finally:
$$
\mepqproutainkrqrp = i \,\int\,\mepqproutjxrp\,\fsqx\,\dfx,
$$
where $\jx \equiv \left({\Box} + m^2\right)\,\vfx$.

The following steps are also similar to the commutative case (see e.g. \cite{12}). We
represent in the same way $\aoutqpr$:
$$
\pqproutl = \pproutovaaoutqprl.
$$
Performing similar calculations (see (\ref{4'_2})) we obtain
\be\label{8_2}
\aoutqpr = - i \,\int\,\left(\dfsqprsy\,\vfy -
\fsqprsy\,\dvfy\right)|_{\xz = + \infty}\,\dty.
\ee
Here we have used the equality
\be\label{9_2}
\int\, [ \; ]|_{\xz = + \infty}\,\dty =
\int\, [ \; ]_{\xz = - \infty}\,\dty +
\int\,\frac{\partial}{\partial\,\yz}\,[ \; ]\,\dfy,
\ee
but in order to show that
$$
\int\, [ \; ]|_{\xz = - \infty}\,\dty = 0
$$
we need some preliminary manipulations with the term on the l.h.s of
(\ref{9_2}). To this end we substitute $\aoutqpr\,\jx$  by $[\aoutqpr, \jx]$. This can be done, since
$$
\mepprjxaoutqprrq = 0.
$$
Then we multiply $[\aoutqpr, \jx]$ by $\tau\,(\yz - \xz)$ as $\yz \to
\infty$. Now the first term in the r.h.s. of (\ref{9_2}) is equal to zero as
$\tau\,(\yz - \xz) = 0$ when $\yz \to - \infty$.

We can also substitute (at $\yz = + \infty$) $\vfy\,\jx$ by $\tau\,(\yz -
\xz)\,\vfy\,\jx + \tau\,(\xz - \yz)\,\jx\,\vfy = T\,(\vfy\,\jx)$. It
remains then to consider the last term in (\ref{9_2}). By calculations similar
to those made above we obtain
\be\label{10_2}
\pqproutqpin = - \int\,\mepprpsxyrp\,\fsqx\,\dfy\,\dfx,
\ee
where
$$
\psxy = \fsqprsy\,\tau\,(\yz - \xz)\,[\jy, \jx] + \deyzmxz\,\left(-
\dfsqprsy\,\vfy + \fsqprsy\,\dvfy\right)
$$
(using $\frac{\partial}{\partial\,\yz}\,\tau\,(\yz - \xz)=\delta(x_0-y_0)$).

Taking into account that
$$
\mepprpsxyrp = \exipprmpa\,\mepprpsxmaymarp,
$$
after translation by $a = \frac {x + y} 2$ and trivial calculations, we
finally obtain:
\bn\label{11_2}
\pqproutqpin &=&- \de\,(\ppr + \qpr - p - q)\cr
&\times&\{\int\fsqprsxh\,\tau\,(\xz)\,< p' \left|
\left[\jxh, \jmxh\right] \right| p >\,\fsqsxh\,\dfx\cr
&+& P\,(q, q')\}.
\en

The last term is some polynomial in $q$ and $q'$. To show this we admit
the standard assumption that $\jx$ is some polynomial in $\vfx$ with $\star$-product, $\theta_{0i}=0$, and use
the equal-time commutation relations.

After integrating over the noncommuting variables $x_1$ and $x_2$ we come to a similar formula with
eq. (\ref{4'}). (We recall that $\theta_{3i}$=0, see the Introduction.) Thus the LSZ reduction formulas are valid in NC case.

\subsection{BMP (Bogoliubov-Medvedev-Polivanov) reduction formulas}

We shall now consider the BMP reduction formulas. In the \co case, the $S$-matrix is
represented in the general form:
\be\label{12_2}
S = \sumzi\,\int\,\fsnxoxn \colon \vfxoxn \colon \dfxo \cdots \dfxn,
\ee
where $\fsnxoxn$ are  some functions, $\vfx \equiv \vfoutx$ and normal
product is used \cite{10,13}.

It is evident that
\bn\label{13_2}
\mepqprsrqp =\mepqprcsakrqrp
\en
since
$$
\mepqprakrqsrp = 0.
$$

We omit here the index "out".

>From (\ref{2_2}) it follows that
\be\label{14_2}
[\vfx, \akrq] = \fsqx.
\ee
Commuting $\akrq$ with $\vfxn, \; \vfxnmo$ and so on, we see that
\be\label{15_2}
[S, \akrq] = \int\,\desdevfx\,\fsqx\,\dx,
\ee
where
\bn\label{16_2}
\desdevfx &\equiv&
\sumzi\,\sumzn\,\int\,\fsnxoxixn\cr
&\times&\colon \vfxo \cdots \widehat{\vfxi} \cdots \vfxn \colon \dfxo \cdots
\widehat{\dfxi} \cdots \dfxn.
\en
The notation $\widehat{\qquad}$ means the absence of the corresponding term.

Thus
\be\label{17_2}
\mepqprsrqp = \int\,\fsqx\,\mepqprdesdevfxrp\,\dfx.
\ee
The next step is similar. We substitute
$$
\meppramqprdesdevfxrp
$$
by
$$
\mepprcamqprdesdevfxrp
$$
and then use the analog of (\ref{15_2})
\be\label{18_2}
[\amqpr, S] =  \int\,\fsqprsy\,\desdevfy\,\dfy.
\ee
In accordance with (\ref{17_2}) and (\ref{18_2})
\bn\label{19_2}
\mepqprsrqp = \int\,\fsqprsy\,\fsqx\,< p' | \dessdevfxdevfy | p >
\,\dfy\,\dfx  \cr
\sim\de\,(\ppr + \qpr - p - q)\,\int\,\exiqpqprxh\,< p' |
\dessdevfxhdevfmxh\,S^* | p >\dfx,
\en
where we have used that $| p > = S^*\,S | p > = S^* | p >$, in accordance with the
unitarity of the $S$-matrix and stability of one-particle state. Eq. (\ref{19_2}) is the BMP
reduction formula.

If we put
$$
\jx \equiv i\,\desdevfx\,S^*, \quad (\jx = \jsx),
$$
we can check that
\be\label{20_2}
\dessdevfxdevfy\,S^* = - T\,(\jx\,\jy)
\ee
and so BMP and LSZ reduction formulas coincide. We should point
out that (\ref{20_2}) follows from the Bogoliubov microcausality condition:
\be\label{21_2}
\dedevfx \,\jy = 0, \quad \mbox{if} \quad \xz <\yz  \quad \mbox{or} \quad
{(x - y)}^2 < 0.
\ee

We turn now towards the \nc case. Here it is natural to represent $S$-matrix by
expression (\ref{12_2}), but using $\star$-product and $f_n\,(\xo, \cdots \xn,
\tod)$ instead of $\fsnxoxn$.

Equality (\ref{14_2}) is valid as before since the descriptions of asymptotic fields
in the \nc and commutative theories are the same when $\theta_{0i}=0$, i.e. time is commutative.
Taking into account that
$$
\int\,f_n\,(\xo, \cdots \xn, \tod)\,\star\,\fsqxn\,\dfxn =
\int\,f_n\,(\xo, \cdots \xn, \tod)\,\fsqxn\,\dfxn
$$
and similar formulas for $\fsqxi$ we come to eq. (\ref{17_2}), where $\desdevfx$
is determined by (\ref{16_2}), but with $\star$-product.

We note that, using integration by parts, we can define $S$-matrix with standard
product, but with redefined $f_n\,(\xo, \cdots \xn, \tod)$.

Eq. (\ref{19_2}) is valid on the same basis as eq. (\ref{17_2}). However, in
equality (\ref{21_2}) we have to substitute the condition ${(x - y)}^2 < 0$ by the
condition $( x_0-y_0 )^2  -( x_3-y_3)^2<0 |$, due to the {\it modified} Bogoliubov microcausality condition
(\ref{28}).


\begin{thebibliography}{99}

\bibitem{1}
M. Gell-Mann, M. L. Goldberger and W. E. Thirring, {\it Phys.
Rev.}  {\bf 95} (1954) 1612.

\bibitem{2}
M. L. Goldberger, {\it Phys. Rev.} {\bf 99} (1955) 979.

\bibitem{3}
M. L. Goldberger, H. Miyazawa and R. Oehme, {\it Phys. Rev.}
{\bf 99} (1955) 986.

\bibitem{4}
R. Oehme, {\it Phys. Rev.} {\bf 100} (1955) 1503; {\bf 102}
(1956) 1174.

\bibitem{5}
N. N. Bogoliubov, {\it Lecture at International Congress on Theoretical
Physics}, Seattle, 1956 (unpublished).

\bibitem{6}
R. Oehme, {\it Nuovo Cim.} {\bf 10} (1958) 1316.

\bibitem{7}
K. Symanzik, {\it Phys. Rev.} {\bf 105} (1957) 743.

\bibitem{8}
H. J. Bremermann, R. Oehme and J.G. Taylor, {\it Phys. Rev.}
{\bf 109} (1958) 2178.

\bibitem{9}
H. Lehmann, {\it Nuovo Cim.}  {\bf 10} (1958) 579.

\bibitem{10}
N. N. Bogoliubov, B. V. Medvedev and M. K. Polivanov, {\it Theory of Dispersion Relations}, Lawrence Radiation
Laboratory, Berkeley, California, 1961.

\bibitem{Connes}
A. Connes, {\it Noncommutative Geometry}, Academic Press, New York, 1994.

\bibitem{Snyder}
H. S. Snyder, {\it Phys. Rev} {\bf 71} (1947) 38.

\bibitem{Dopli}

S. Doplicher, K. Fredenhagen and J. E. Roberts, {\it Phys. Lett} {\bf
B331} (1994) 39; {\it Comm. Math. Phys.} {\bf 172} (1995) 187.

\bibitem{Shahin}

F. Ardalan, H. Arfaei and M. M. Sheikh-Jabbari, {\it JHEP} {\bf 9902} (1999) 016, hep-th/9810072.

\bibitem{SW}
N. Seiberg and E. Witten, {\it JHEP} {\bf 9909} (1999) 32, hep-th/9908142
and references therein.

\bibitem{DN}
M. R. Douglas and N. A. Nekrasov, {\it Rev. Mod. Phys.} {\bf 73} (2001) 977, hep-th/0106048.

\bibitem{LAG}
L. Alvarez-Gaum\'e, J. L. F. Barbon and R. Zwicky, {\it JHEP} {\bf 0105} (2001) 057, hep-th/0103069.

\bibitem{17}
M. Chaichian, K. Nishijima and A. Tureanu, hep-th/0209006, to appear in Phys. Lett. B.

\bibitem{SST}
N. Seiberg, L. Susskind and N. Toumbas, {\it JHEP} {\bf 0006} (2000) 044, hep-th/0005015.

\bibitem{11}
Y. Liao and K. Sibold, {\it Phys. Lett.} {\bf B 549} (2002) 352, hep-th/0209221.

\bibitem{LSZ}
H. Lehmann, K. Symanzik and W. Zimmermann, {\it Nuovo Cim.} {\bf 1} (1955) 1425; {\bf 6} (1957) 319.

%
\bibitem{12}
J. D. Bjorken, S. D. Drell, {\it Relativistic Quantum Fields}, Mc Graw-Hill
Book Company, 1965.

\bibitem{23ad}
R. Streater and A.S. Wightman, {\it PCT, Spin
and Statistics, and All That}, Benjamin, New York, 1964.
\bibitem{13}
N. N. Bogoliubov and D. V. Shirkov, {\it Introduction to the Theory of Quantized
Fields}, Wiley, New York, 1980, 3rd ed.

\bibitem{Froissart}
M. Froissart, {\it Phys. Rev.} {\bf 123} (1961) 1053.

\bibitem{Martin}
A. Martin, {\it Phys. Rev.} {\bf 129} (1963) 1432.

\bibitem{Martin2}
A. Martin, {\it Nuovo Cim.} {\bf 42} (1966) 901.

\bibitem{14}
N. N. Meiman, {\it Zh. Eksp. Teor. Fiz.} {\bf 43} (1962) 2247 (translation in
{\it Sov. Phys. JETF} {\bf 16} (1963) 1609).

%

\bibitem{18}
V. Ya. Fainberg and Sh. Yu. Lomsadze, {\it Kratkie
Soobshcheniya po Fizike} {\bf 5} (1988) 23 (translated in {\it Soviet Physics - Lebedev Institute
Reports}).

\bibitem{19}
Yu. S. Vernov and M. N. Mnatsakanova, in {\it Proceedings of the XIV
International Seminar on High Energy Physics and Quantum Field Theory},
Protvino, Nauka, 1992, p. 290.

\bibitem{CMT}
M. Chaichian, C. Montonen and A. Tureanu, hep-th/0305243.



\end{thebibliography}
\end{document}